\documentclass[a4paper,12pt]{article}
\usepackage{graphicx}
\usepackage{amsmath}
\usepackage{multirow}
\usepackage{float} 
\usepackage{url,tabularx,amsfonts,slashed,array,mcite}
\begin{document}

\begin{flushright}
preprint SHEP-11-12\\
\today
\end{flushright}
\vspace*{1.0truecm}

\begin{center}
{\large\bf Top quark phenomenology of the ADD model\\[0.2cm]
and the Minimal Length Scenario}\\
\vspace*{1.0truecm}
{\large K. Mimasu and S. Moretti}\\
\vspace*{0.5truecm}
{\it School of Physics \& Astronomy, \\
 University of Southampton, Southampton, SO17 1BJ, UK}
\end{center}

\vspace*{1.0truecm}
\begin{center}
\begin{abstract}
\noindent 
We study top-(anti)quark pair production at the Tevatron and Large Hadron Collider (LHC)
in the context of the Minimal Length Scenario (MLS) of the Arkani-Hamed, Dimopoulos and Dvali (ADD)
 model of extra dimensions (XDs). We show that 
sizable effects onto both the integrated and differential cross section due to graviton mediation
are expected for a String scale, $M_S$, of ${\cal O}(1-10~{\rm TeV})$
and several XDs, $\delta$, all compatible with current experimental constraints. Potential 
limits on $M_S$ are extracted. We also highlight clear phenomenological differences between 
a simple ADD scenario and its modification based on using the MLS as a natural regulator for divergent amplitudes of virtual KK graviton exchange.
\end{abstract}
\end{center}

\section{Introduction}
\label{sect:intro}
\noindent
Models of extra dimensions (XDs) have emerged in the past decade or so as an alternative approach to physics 
Beyond the Standard Model (BSM). Although such notions were conceived almost a 
century ago, it took until the advent of String theory to place them on a sound theoretical footing, 
culminating in the proposal by Arkani-Hamed, Dimopoulos and Dvali~\cite{arkani1998hierarchy} of large XDs. 
Known as the ADD model, it reformulated the gauge hierarchy problem of the Standard Model (SM) in terms of
compact extra dimensions in which only gravity can propagate. This dilutes the strength of gravity 
relative to other forces and reconciles the observed Planck mass, $M_{P}$, with the 
Electro-Weak (EW) scale via a new fundamental string scale, $M_{S}$, associated with gravity in the 
bulk. The relatively large volume of such XDs can lower the effective gravity scale to $\mathcal{O}(\text{TeV})$, 
which leads to the phenomenologically interesting case where 
gravitational effects could be observed at modern collider experiments. The relationship between the two scales 
(Planck and String) in terms of the volume of $\delta$ extra spatial dimensions goes as
\begin{align}\label{eq:hierarchy}
	M_{P}^{2} = V_{\delta}M_{S}^{2+\delta}.
\end{align}
All SM degrees of freedom are confined to a 3-brane embedded in the bulk, a scenario that can be realised 
in String theory~\cite{AntoniadisPhys.Lett.B436:257-2631998}. Such models can be understood as low energy effective realisations of String type 
theories where bulk gravitons emerge in 4 dimensions as infinite towers of Kaluza-Klein (KK) modes coupling to the SM with a strength suppressed by $M_{P}$. 
Accounting for the sum over all of these modes, however, leads to an effective coupling of order $1/M_{S}$. 
Therefore, one could observe effects due to real and virtual production of KK gravitons at the 
Tevatron and the Large Hadron Collider (LHC) through exotic missing energy signals and anomalous angular distributions as well as 
statistically significant deviations from those SM cross sections which are known precisely.\\

\noindent
Another concept that arises from String theory is that of a minimum length scale below which one cannot 
probe. In short, if the energy of a String reaches $M_{S}$, perturbative String theory tells us that 
excitations would cause its extension~\cite{witten:24}. The aforementioned possibility that the String scale 
can be lowered to experimentally accessible energies allows for a minimal length, $l_{S}(\sim 1/M_{S})$, 
to also have phenomenological consequences. In particular, it can provide a mechanism to smoothly regularise 
Ultra-Violet (UV) divergences that occur in the amplitudes for virtual KK graviton 
production~\cite{Hossenfelder:2003jz}.\\

\noindent
We describe a study of the ADD model extended with this phenomenological interpretation of a 
Minimum Length Scenario (MLS) in the context of virtual graviton exchange in the $t\bar{t}$ channel at current hadron 
collider experiments. The aim is to determine the reach of the Tevatron and the LHC to set bounds on 
$M_{S}$ by considering effects on the total $pp(\bar{p})\rightarrow t\bar{t}$ production cross section as well
as various differential distributions due to the ADD model and its MLS extension. This is inspired by a similar 
study conducted on the Drell-Yan (DY) channel in~\cite{Bhattacharyya:2004dy}. While the di-lepton channel is the most common 
probe of the ADD model, other studies considering $t\bar{t}$ in the context of the ADD model have also focused on the 
spin properties of KK gravitons to extract spin correlations~\cite{AraiPhys.Rev.D70:1150152004}. As will be discussed later, the 
$t\bar{t}$ channel becomes important at the LHC due to the dominance of gluon initial states, especially in 
the context of the ADD model where interference terms with the SM in the $gg$ channel are the main contribution 
to observable deviations.\\

\noindent
This paper is organised as follows. Sec.~\ref{sect:ADDMLS} goes into some more details of the ADD model and the 
consequences of its MLS extension. In Sec.~\ref{sect:ppttbar}, we consider the model in terms of $t\bar{t}$ production at hadron colliders. 
Sec. \ref{sect:res} presents our results. Finally, we summarise and conclude in Sec. \ref{sect:summa}.\\

\clearpage
\section{Large XDs and the MLS}
\label{sect:ADDMLS}
\subsection{The ADD model} 
\label{sub:ADD}
\noindent
The ADD model represents a simple realisation of $\delta$ transverse extra dimensions, where the SM is fixed to 
exist on a 3-brane embedded in the bulk and the fundamental scale, $M_{S}$, is low ($\mathcal{O}(\text{TeV})$). 
The bulk has the structure of a product space of 4-dimensional (4D) Minkowski space and a $\delta$ dimensional compact
subspace, 
chosen to be a torus with a common compactification radius, $R$, for simplicity ($M_{4}\times T_{\delta}$). 
This amounts to a model of linearised gravity in $D=4+\delta$ dimensions where the degrees of freedom arise 
from fluctuations of the induced metric on the brane about flatness. The D dimensional graviton breaks down into a 4D graviton, $\delta$ vectors and $\delta(\delta+1)/2$ scalars -- corresponding to the zero modes of the KK spectrum. Above these lie infinite towers of closely spaced KK modes with masses $m_{n} = 2\pi \vert\vec{n}\vert/R$, {where} $\vec{n}=(n_{1},n_{2},\dots,n_{\delta})$, with $n_i$ labelling the KK number in each extra dimensional direction. After fixing the residual gauge freedom from general coordinate transformations one is left with a massive graviton, $\delta-1$ vectors and $\delta(\delta-1)/2$ scalars~\cite{Csaki2007} per KK level.\\ 

\noindent
This study follows the definitions and conventions established in~\cite{Han:1998sg} where KK decomposition 
shows that the vector and all but one scalar decouple from the SM while the gravitons, $h^{\vec{n}}_{\mu\nu}$, 
couple universally to the energy-momentum tensor. The remaining scalar degree of freedom, $\phi^{\vec{n}}$, couples to the trace of the energy-momentum tensor. This is the `radion' associated with 
fluctuations of the size of the XDs. The effective interaction Lagrangian in 4D 
can be shown to be~\cite{Han:1998sg}:
\begin{equation}
	\mathcal{L}_{int}= -\frac{\kappa}{2}(h^{\vec{n}}_{\mu\nu}T^{\mu\nu} + \phi^{\vec{n}}T^{\mu}_{\mu}),\quad \text{where}\quad
	T_{\mu\nu}\equiv-\eta_{\mu\nu}\mathcal{L}_{SM} + 2\frac{\delta\mathcal{L}_{SM}}{\delta g^{\mu\nu}}\Big\vert_{g_{\mu\nu}=\eta_{\mu\nu}}
\end{equation}
and the coupling constant is defined as $\kappa^{2} = 16\pi/M_{P}^{2}$.\\

\noindent
The fact that the SM is localised on the brane means that, in the limit of a weak gravitational field, one can consider the energy-momentum tensor to have 
KK modes independent of $n$, which provides the universal coupling of all graviton excitations in energy regimes 
lower than the cutoff $M_{S}$, as discussed in~\cite{Giudice:1998ck}. The effects due to bulk deformations of the SM brane are not considered here in that its tension is 
assumed to be very large ($\gg M_{S}$) and hence the `branons' very heavy. 

\subsection{Virtual gravitons and the MLS} 
\label{sub:virtualMLS}
\noindent
The Feynman rules for the vertices coupling the graviton to the 
SM are summarised in~\cite{Han:1998sg}; from these, one can calculate Matrix Elements (MEs) and hence cross sections for 
processes such as the one of interest: $pp(\bar{p})\rightarrow t\bar{t}$. In the SM this is dominated by 
Quantum Chromo-Dynamics (QCD) with a tiny EW contribution while in the ADD model it is mediated by an $s$-channel virtual 
KK graviton. Since the initial state particles can be assumed to be massless, the contribution from the scalar 
graviton sector is negligible due to its coupling to the trace of $T_{\mu\nu}$. What remains is to sum up the 
contributions from all KK modes. This is dealt with in the appendix of~\cite{Han:1998sg} by treating the sum over 
many propagators as an integral over the KK mode mass:
\begin{equation}
	D(p^{2}) = \sum_{\vec{n}}\frac{1}{p^{2}-m_{n}^{2}+i\epsilon} \Rightarrow \int^{\Lambda^{2}}\frac{\rho(m_{n})dm_{n}^{2}}{p^{2}-m_{n}^{2}+i\epsilon},
\end{equation}
where the tensor structure of the numerator has been dropped. The density function, $\rho(m_{n})$, 
accounts for degenerate KK levels. The tree-level amplitude for the mediation of a virtual KK graviton gains 
an effective coupling factor, $\lambda_{\rm{eff}}$, from this integral:
\begin{equation}\label{eq:addint}
	\lambda_{\rm{eff}}\sim D(s)\kappa^{2} \sim \frac{F(\hat{s},\delta)}{M_{S}^{4}}P\left[I\left(\tfrac{M_{S}}{\sqrt{\hat{s}}}\right)\right];\quad I\left(\tfrac{M_{S}}{\sqrt{\hat{s}}}\right)=\int^{M_{S}/\sqrt{\hat{s}}}_{0}dy\frac{y^{\delta-1}}{1-y^{2}},
\end{equation}
where $F$ is a function of the partonic centre of mass (CM) energy, $\hat{s}$, and the number of XDs while $P$ denotes the principal part 
of the integral. A hard cutoff associated with $M_{S}$ must be introduced due to the divergent
nature of this integral.\\

\noindent
The MLS -- defined as the incorporation of a minimum length, $l_{S}$, into the ADD setup -- provides an alternative 
mechanism for regularising these divergences. If one assumes this scenario the uncertainty in position 
measurement can no longer be less than $l_{S}$. This entails a modification of the relationship between 
momentum $p$ and wave vector $k$ such that, whilst particle momentum can become arbitrarily high, the wave vector 
is bounded by $1/l_{S}$. A simple parametrisation that captures the essence of a 
MLS is that of the Unruh relations~\cite{PhysRevD.51.2827}:
\begin{equation}
\begin{split}
	k(p)&=\frac{1}{l_{S}}\text{tanh}^{1/\gamma}\left[\left(\frac{p}{M_{S}}\right)^{\gamma}\right],\\
	\omega(E)&=\frac{1}{l_{S}}\text{tanh}^{1/\gamma}\left[\left(\frac{E}{M_{S}}\right)^{\gamma}\right],
\end{split}
\end{equation} 
where $\gamma$ is henceforth assumed to be 1 for simplicity\footnote{Different choices of $\gamma$ simply would correspond to alternative ways
of reaching $1/l_{S}$, while the asymptotic behaviour would remain the same. The limit $\gamma \rightarrow \infty$ tends toward recovering a hard cutoff at $M_{S}$.}. As discussed in~\cite{Hossenfelder:2003jz}, this scenario brings about a number of modifications to the calculations. Most importantly, the new relations modulate the KK mode mass integral by a factor $\partial \omega/\partial E$ to smoothly cut off the integral, which 
now becomes~\cite{Bhattacharyya:2004dy}
\begin{equation}\label{eq:mlsint}
	I^{\prime}=\int^{\infty}_{0}dy\frac{y^{\delta-1}}{1-y^{2}}\text{sech}^{2}\left(\frac{\sqrt{\hat{s}}}{M_{S}}y\right).
\end{equation}
The integral is finite and can be evaluated numerically. Thus the requirement of a minimal length removes the need for a hard 
energy cutoff.\\

\noindent
Another important modification due to this assumption affects the phase space measure, which is shown 
in~\cite{Hossenfelder:2003jz} to lead to a change in the expression for the fully differential cross section: 
\begin{equation}
	d\tilde{\sigma}=d\sigma\prod_{n}\frac{E_{n}}{\omega_{n}}\prod_{\nu}\frac{\partial k_{\nu}}{\partial p_{\nu}}.
\end{equation}
It is important to stress the phenomenological interpretation of the MLS that is applied in this study. We consider it as a useful mechanism for regulating the KK mass integrals rather than a concrete statement on the geometry of spacetime or the physics beyond $M_{S}$. For example, an issue that lies beyond the scope of this work is that of Lorentz symmetry violation, briefly discussed 
in~\cite{Hossenfelder:2003jz}, coming from the non-linear relationship between $p$ and $k$. The MLS is treated here 
as a mechanism for regularising integrals whereas various other works elaborate in a more rigorous way upon notions of 
corrections to the uncertainty relations as well as of modifications to the Lorentz group action in the context of a 
minimum length or maximum momentum scale~\cite{PhysRevD.52.1108,*PhysRevD.55.7909,*Magueijo2002,*Magueijo2003,*Mu2009}.

\section{The $pp(\bar{p})\rightarrow t\bar{t}$ process in the ADD model} 
\label{sect:ppttbar}
\begin{figure}[h!]
\begin{center}
\includegraphics[width=0.9\linewidth]{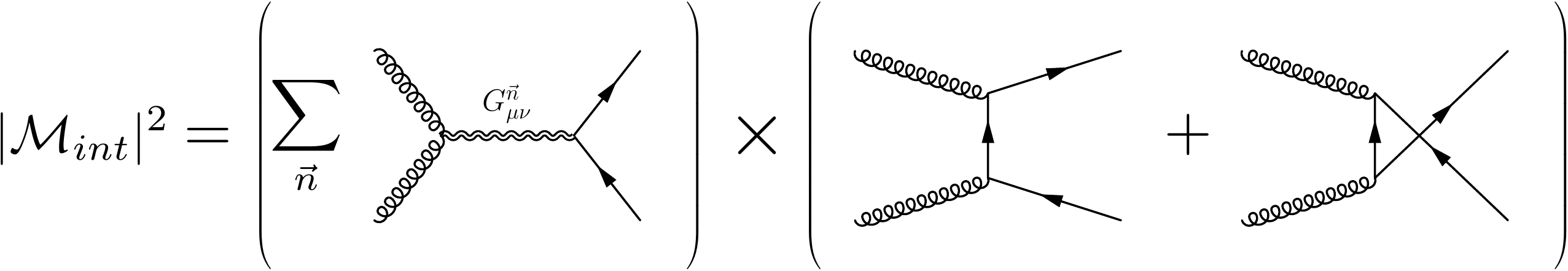}
\caption{ Feynman diagrams showing schematically the interference between tree-level contributions to 
$gg\rightarrow t\bar{t}$ of ADD and QCD.}
\label{fig:int}
\end{center}
\end{figure}

\noindent
A typical cross section for a SM process including ADD effects will take the form:
\begin{equation}
\sigma_{\rm tot} = 
\sigma_{\rm SM} + \sigma_{\rm int}\big[\mathcal{O}(\lambda_{\rm{eff}})\big] + \sigma_{\rm ADD}\big[\mathcal{O}(\lambda_{\rm{eff}}^{2})\big], 
\end{equation}
where $\sigma_{\rm int}$ and $\sigma_{\rm ADD}$ denote the interference and pure ADD contributions, respectively and $\lambda_{\rm{eff}}$ is defined as in eqn.~\ref{eq:addint}. For the QCD 
dominated $t\bar{t}$ production, the $q\bar q$ initial state is mediated by a gluon, meaning that the colour structure of the 
QCD diagram prevents interference with the ADD one. Through the $gg$ initial state, in contrast, 
the SM diagram can interfere with the ADD one in the $t$ and $u$-channels as shown in figure~\ref{fig:int}.\\

\vskip-5mm
\noindent
One would therefore expect the LHC to be a particularly suitable environment to probe 
the ADD model and set bounds on its 
parameters, indeed more than the (mostly $q\bar q$ mediated) Tevatron. This lends weight to the use of the $t\bar{t}$ channel as a worthwhile addition to the various other analyses aiming to constrain large extra dimensions. The amplitudes for the processes 
$gg\rightarrow t\bar{t}$ and $q\bar{q}\rightarrow t\bar{t}$ have been published 
in~\cite{Mathews:1998kf}, including both the interference and pure ADD terms.
The squared amplitudes due to pure QCD were taken from MadGraph \cite{Stelzer:1994ta}. EW processes 
were also taken into account, but merely for completeness, as their effect was practically negligible.\\

\noindent
Based on these results, we used Monte Carlo methods to generate invariant mass distributions and total cross
 sections for the ADD model with and without the MLS over a region of ($M_{S}$, $\delta$) parameter space at three 
benchmark collider setups: LHC at 14 TeV, Tevatron and `early data' LHC at 7 TeV. In each case we assumed an integrated 
luminosity of 100, 5 and 4 fb$^{-1}$, respectively. By comparing these data to SM predictions, the aim is to 
make statements about the reach of the experiments and their capacity to extract XD effects or else set bounds on the 
corresponding parameters. Most results presented here exploited the invariant mass of the final state, $M_{t t}$.
In addition, comparisons in the so called `centrality ratio', $\chi = \exp(\vert y_{1}-y_{2}\vert)$, were also considered, 
where $y_{1,2}$ are the rapidities of the two final state top (anti)quarks. A recent study on early LHC data~\cite{Franceschini2011} used an effective operator approach to ADD type models to extract powerful bounds on the effective coupling of the main operators that contribute to the dijet cross section using this variable. This prompted us to include $\chi$ in our study. The translation of the bounds obtained from this paper to bounds on the fundamental scale of a specific model of large XD's depends on how the KK graviton sum is regulated. The reach obtained in this study extends beyond current experimental constraints and we argue in the next section that a smooth cutoff prescription along with the phase space corrections brought about by the MLS lessen the risk of overestimating ADD contributions at energies near the cutoff scale in lieu of a full understanding of the high energy theory.

\section{Results}
\label{sect:res}
\noindent
A numerical routine was used to calculate the total 
cross sections integrated over a range of invariant mass, $M_{tt},$ for a choice of values of 
$M_{S}$ in both the ADD and ADD-MLS models and compared to the SM, taking into account both the gluon and quark initial states. 
The corrections to the effective coupling and the phase space coming from the MLS were implemented with the 
guidance of~\cite{Hossenfelder:2003jz}, as described in section~\ref{sect:ADDMLS}. 
The CTEQ6 Parton Distribution Functions (PDFs) \cite{cteq} were used to compute the hadronic cross sections, with $Q^2=(2 m_t)^2$.\\
\noindent
No higher order effects were considered, neither in the SM nor for the ADD model in order to consistently compare tree level predictions. This is done with the confidence that the cross section is well known to 
Next-to-Leading Order (NLO) accuracy in the full SM
\cite{Nason:1987xz,*Beenakker:1990maa,*Bernreuther:2004jv,*Kidonakis2008,*Cacciari2008,*Czakon2008,*Melnikov2009,*Bredenstein2010,*Moretti:2006nf, *KuhnEur.Phys.J.C51:37-532007,*Hollik2007,*PhysRevD.74.113005} and therefore normalisation to the SM background is not an issue, 
rather we focus on the potential to observe deviations from predictions due to the ADD and the MLS dynamics. All
data shown (except in subsecs.~\ref{sub:total} and~\ref{sub:lumi}) correspond to the LHC at 14 TeV
as, in general, the other benchmark collider setups neither show qualitatively different results nor offer
better sensitivity. 
\subsection{Differential cross sections}
\label{sub:dists}
\subsubsection{Invariant mass}
\label{sub:mttdists}
\begin{figure}[!h]
\begin{center}
\includegraphics[width=0.49\linewidth]{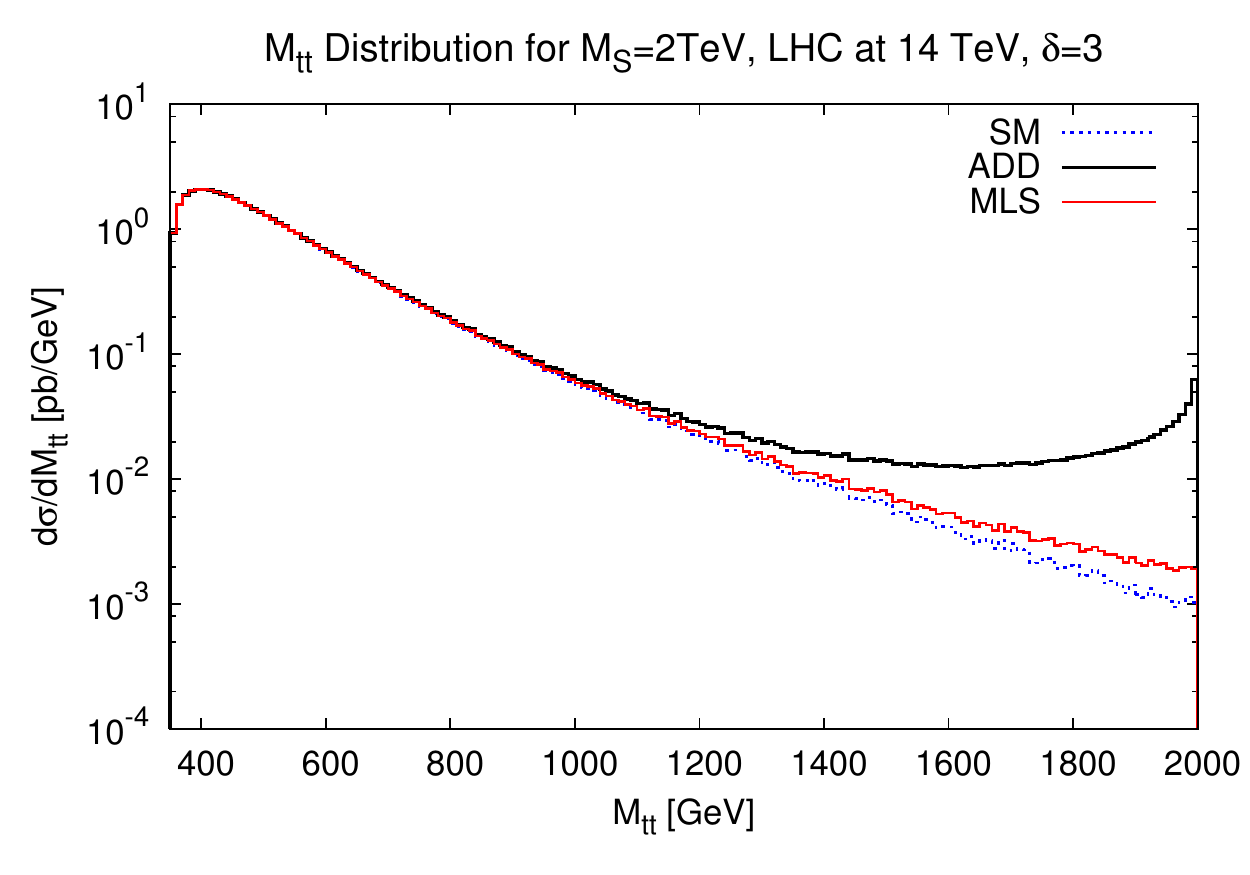}
\includegraphics[width=0.49\linewidth]{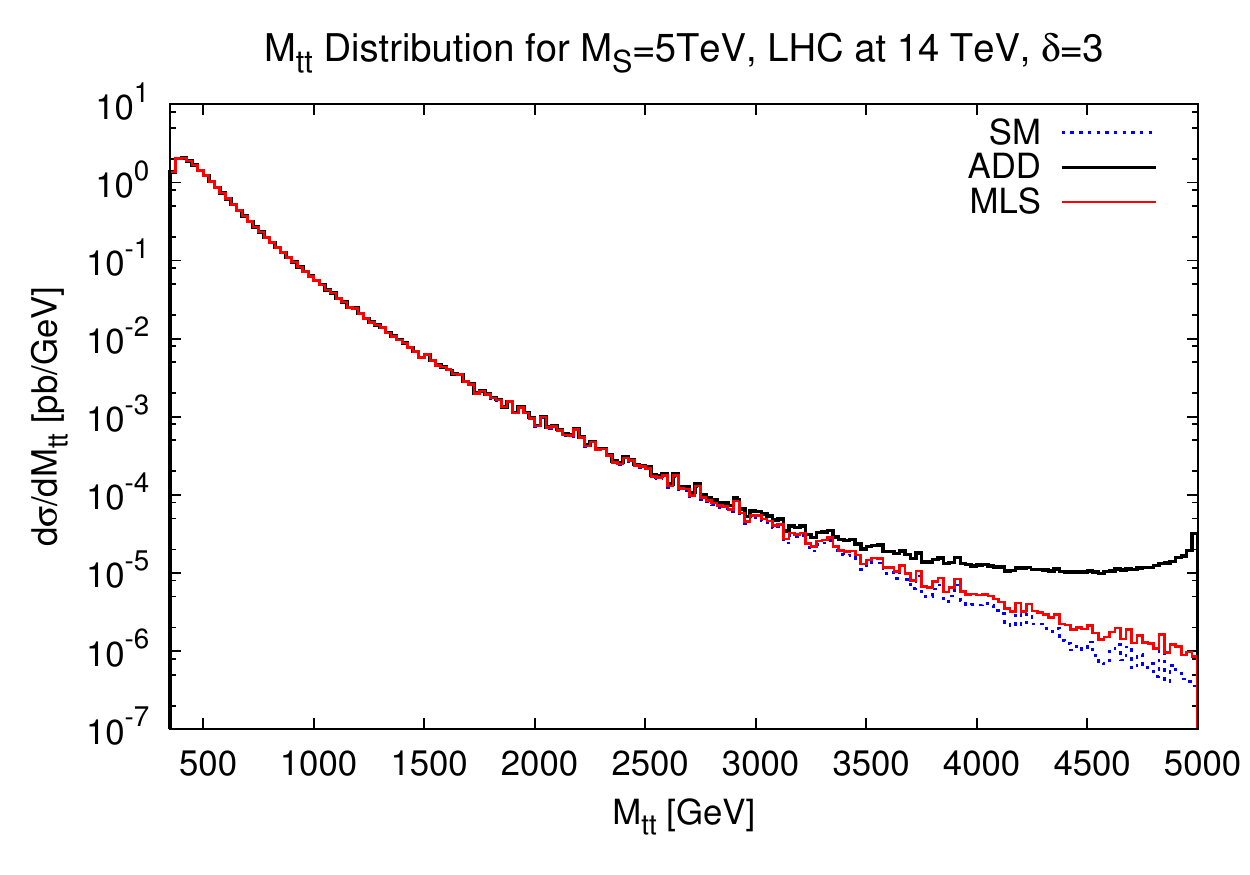}
\includegraphics[width=0.49\linewidth]{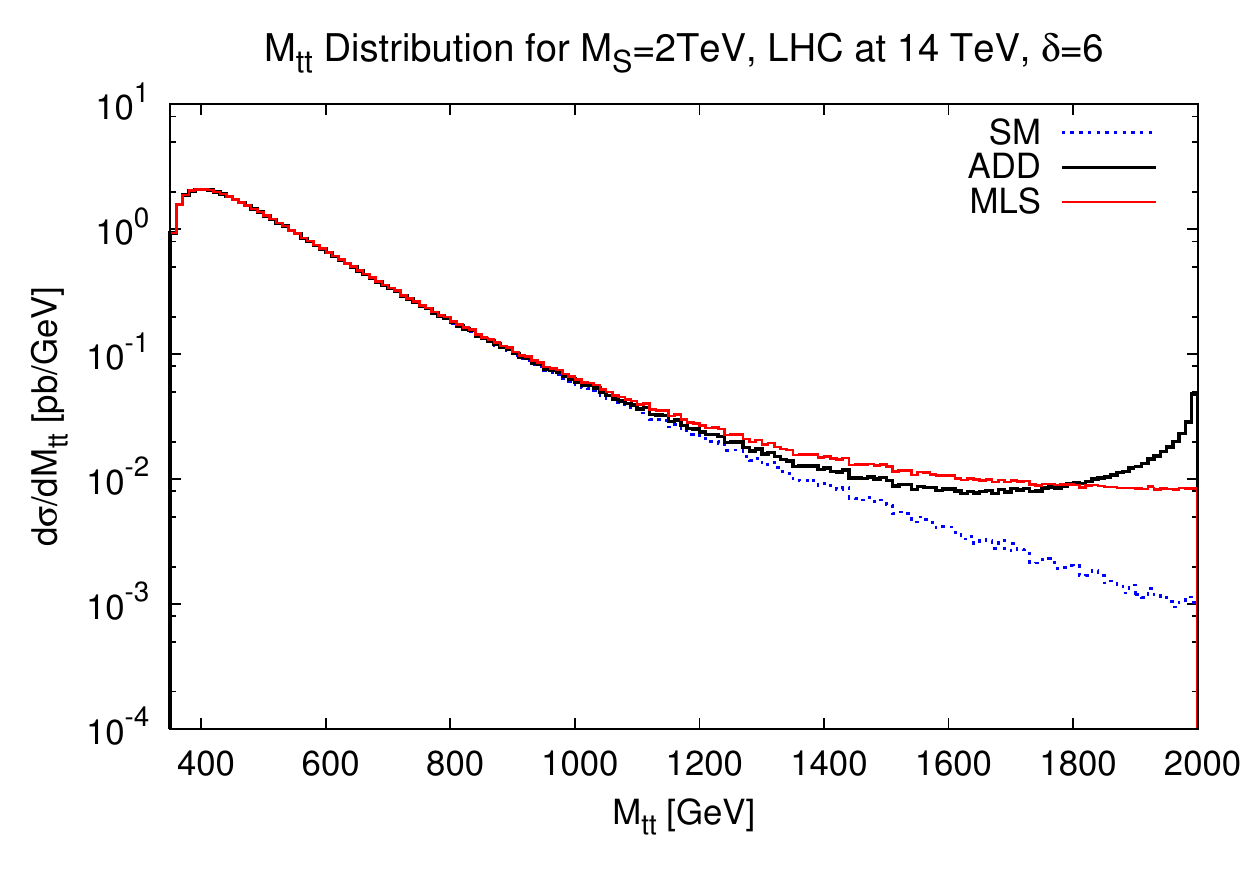}
\includegraphics[width=0.49\linewidth]{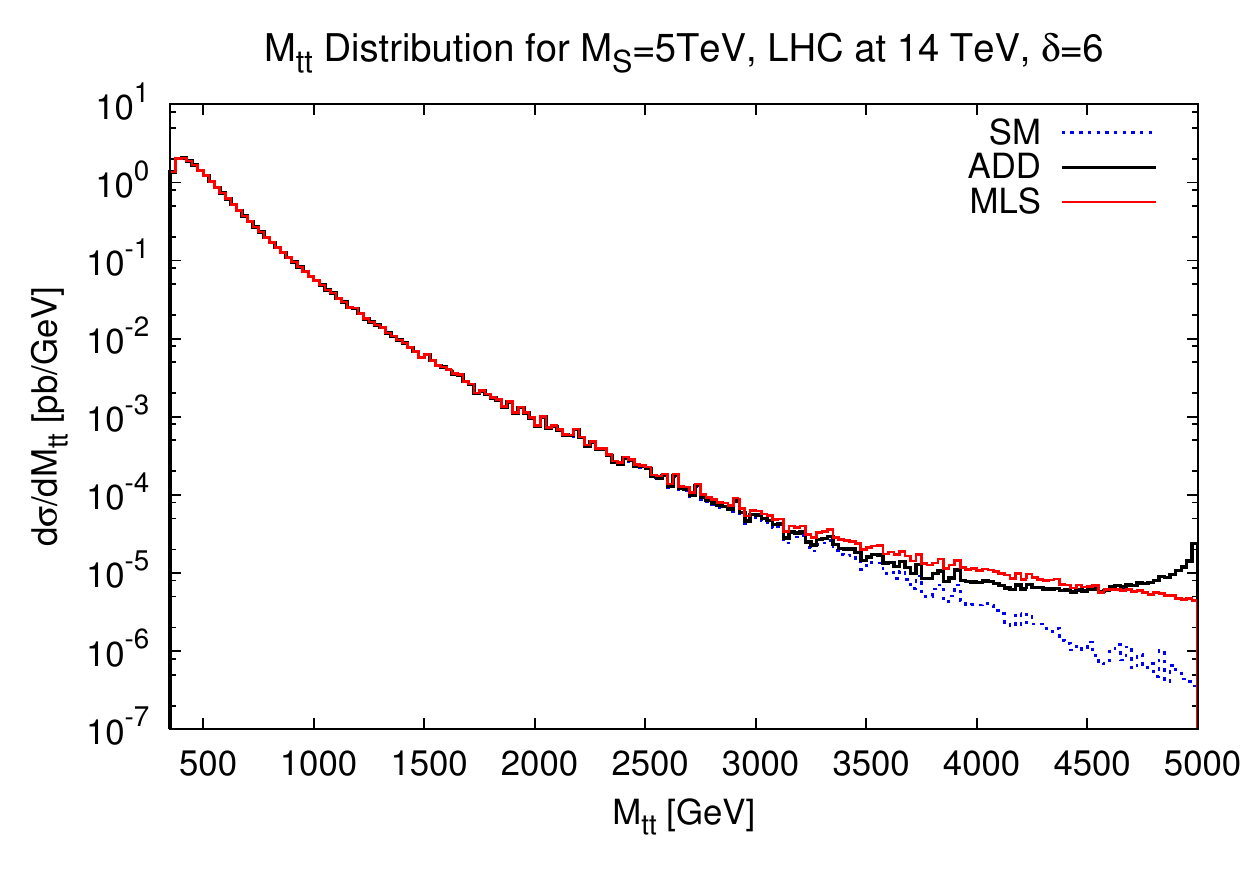}
\centering
\caption{ Differential cross sections w.r.t. the $t\bar{t}$ invariant mass, 
$M_{tt}$, at the LHC at 14 TeV for $M_{S}$~=~2 and 5 for the SM, ADD and ADD-MLS 
for $\delta~=~3(6)$ XDs [\emph{upper}(\emph{lower})].}
\label{fig:lhc14mttdists}
\end{center}
\end{figure}

\noindent
We start our analysis by looking at the differential cross section in invariant mass of the
top-antitop pair, $M_{tt}$, for the purpose of determining 
the ranges over which to integrate upon in order to extract good signal-to-background ratios. 
This also provides information about
energy regimes in which the validity of such effective theories may be called into question. 
Unlike the total cross section, invariant mass distributions give some insight into the behaviour 
near the cutoff, $M_{S}$, where one would expect the effective model to start breaking down. 
This is reflected in Fig.~\ref{fig:lhc14mttdists} where the ADD cross section appears in fact to 
diverge near $M_{S}$ and where we can also highlight the naturally regulating 
feature of its MLS extension, which does not have this asymptotic behaviour.\\

\noindent
The absence of such divergence in the MLS case shows that the $\sim {\rm sech}^{2}(\frac{\sqrt{s}}{M_{S}})$ 
suppression in the mass integral works in taming the effective coupling while the phase space factors also 
further deplete the cross section at high invariant mass. This leads to the conclusion that a valid comparison of 
all models can occur only if one does not integrate up to $M_{S}$ but up to somewhere like $0.8M_{S}$, where one starts 
to see an unnatural enhancement of the ADD cross section, thus avoiding an over-estimation of the ADD contribution. We find that it may be difficult 
to produce any observable deviation from the SM at very high $M_{S}$, especially when it is above the collider energy.\\

\begin{figure}[!h]
\begin{center}
\includegraphics[width=0.49\linewidth]{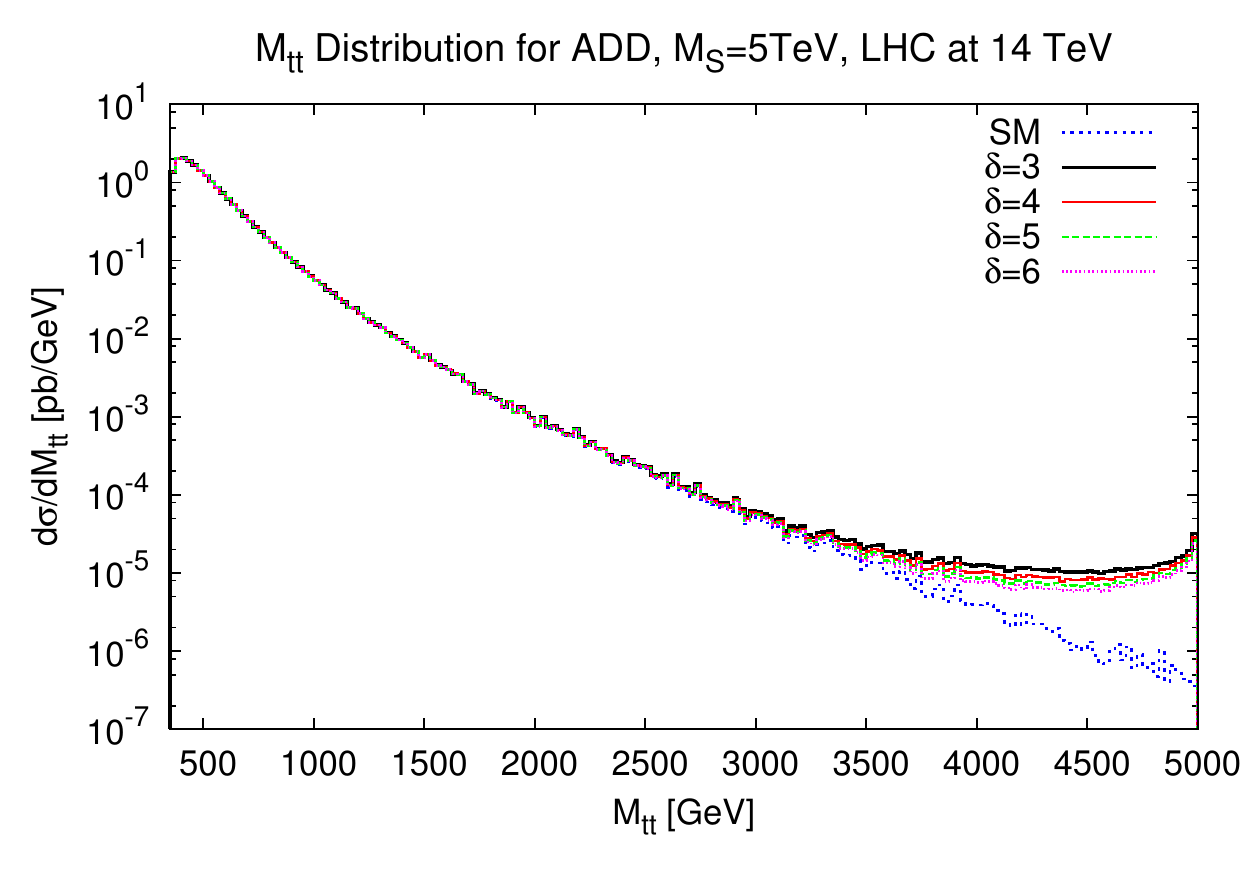}
\includegraphics[width=0.49\linewidth]{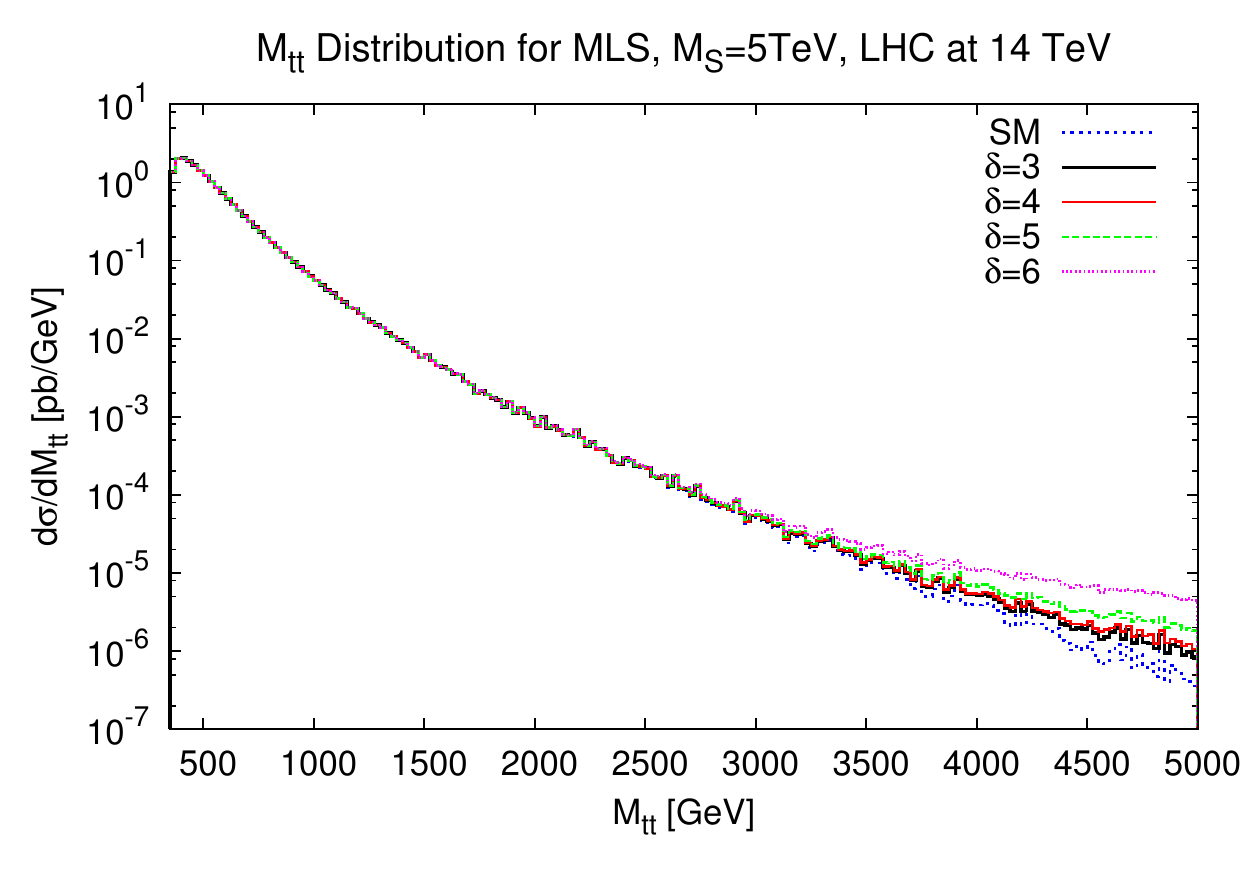}
\caption{ Variation with the number of XDs, $\delta$, of the differential cross sections 
w.r.t. the $t\bar{t}$ invariant mass, $M_{tt}$, for the LHC at 14 TeV and $M_{S}$~=~5 TeV shown in the ADD and 
ADD-MLS scenarios.}
\label{fig:lhc14mttdistsN}
\end{center}
\end{figure}

\noindent
The figures also provide guidance on the minimum invariant mass cuts needed for the eventual total cross section 
calculations. A `tracking' invariant mass cut of $M_{tt}>M_{S}/2$ was decided upon as a simple and effective one to implement. It is interesting to note that, over most of the range where deviation from the SM 
differential cross section occurs, the MLS spectrum becomes more and more comparable to the ADD one 
as the number of XDs is increased, surpassing it at $\delta=6$. This is a consequence of the fact 
that the MLS allows one to integrate over the whole range of KK modes while the ADD restricts the process to the 
exchange of modes up to the effective cutoff, $M_{S}$ (compare eqns.~\ref{eq:addint} and~\ref{eq:mlsint}). Fig.~\ref{fig:lhc14mttdistsN} highlights the behaviour of the two models with varying $\delta$.
\subsubsection{Centrality ratio}
\label{sub:chidists}
\noindent
Another useful differential observable is the
centrality ratio, $\chi$, defined as $\exp(\vert y_{1}-y_{2}\vert)\equiv\exp(\vert2y^{\ast}\vert)$, 
where $y_{1,2}$ are the rapidities of the final state particles in the collider frame and $y^{\ast}$ is the rapidity 
in the CM frame. In eliminating the boost of the collider frame, it 
allows for a better comparison of angular distributions. The ADD and MLS processes 
proceed via the exchange of a spin-2 graviton which involves terms with a different ($\sim\cos^{4}{\theta}$) angular 
dependence to the SM background, making $\chi$ an effective and robust discriminating variable.\\

\begin{figure}[!h]
\begin{center}
\includegraphics[width=0.49\linewidth]{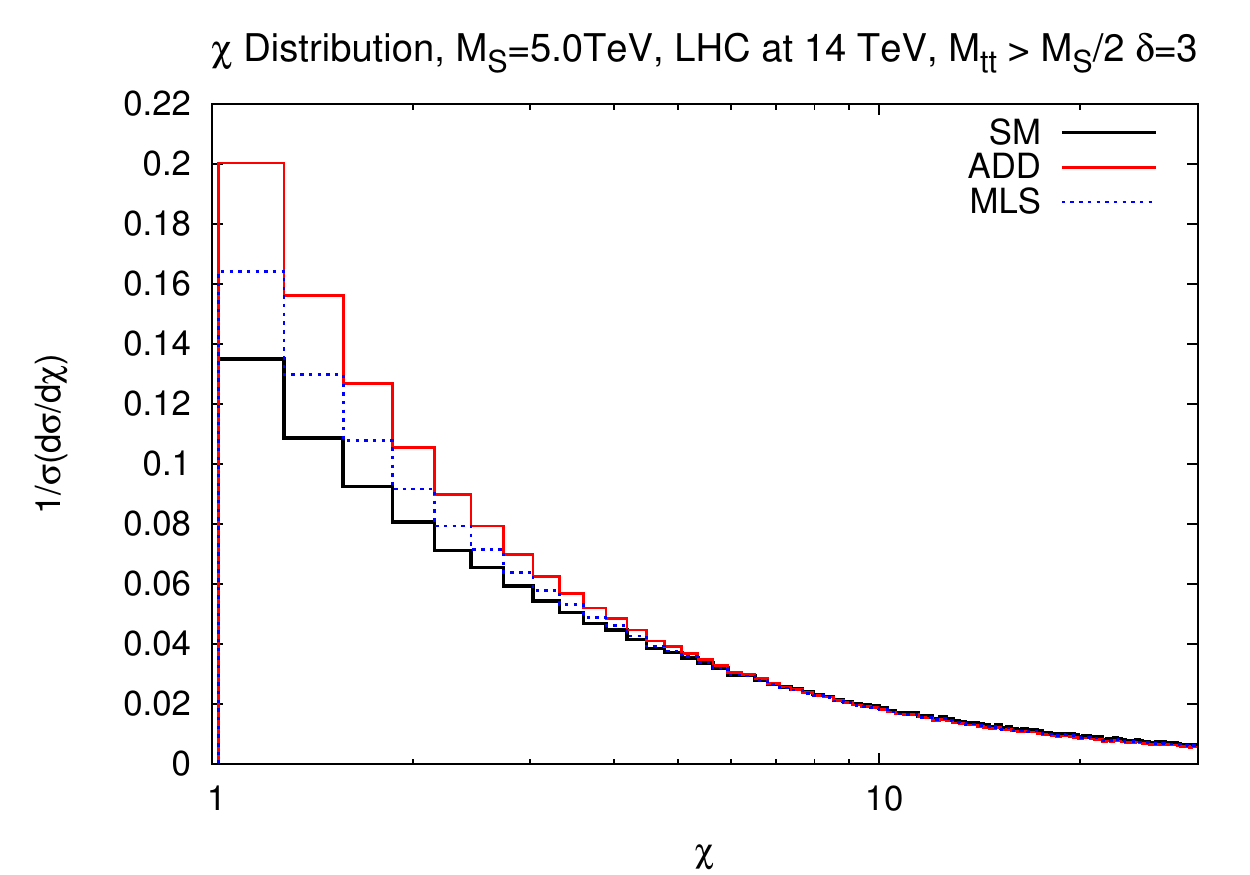}
\includegraphics[width=0.49\linewidth]{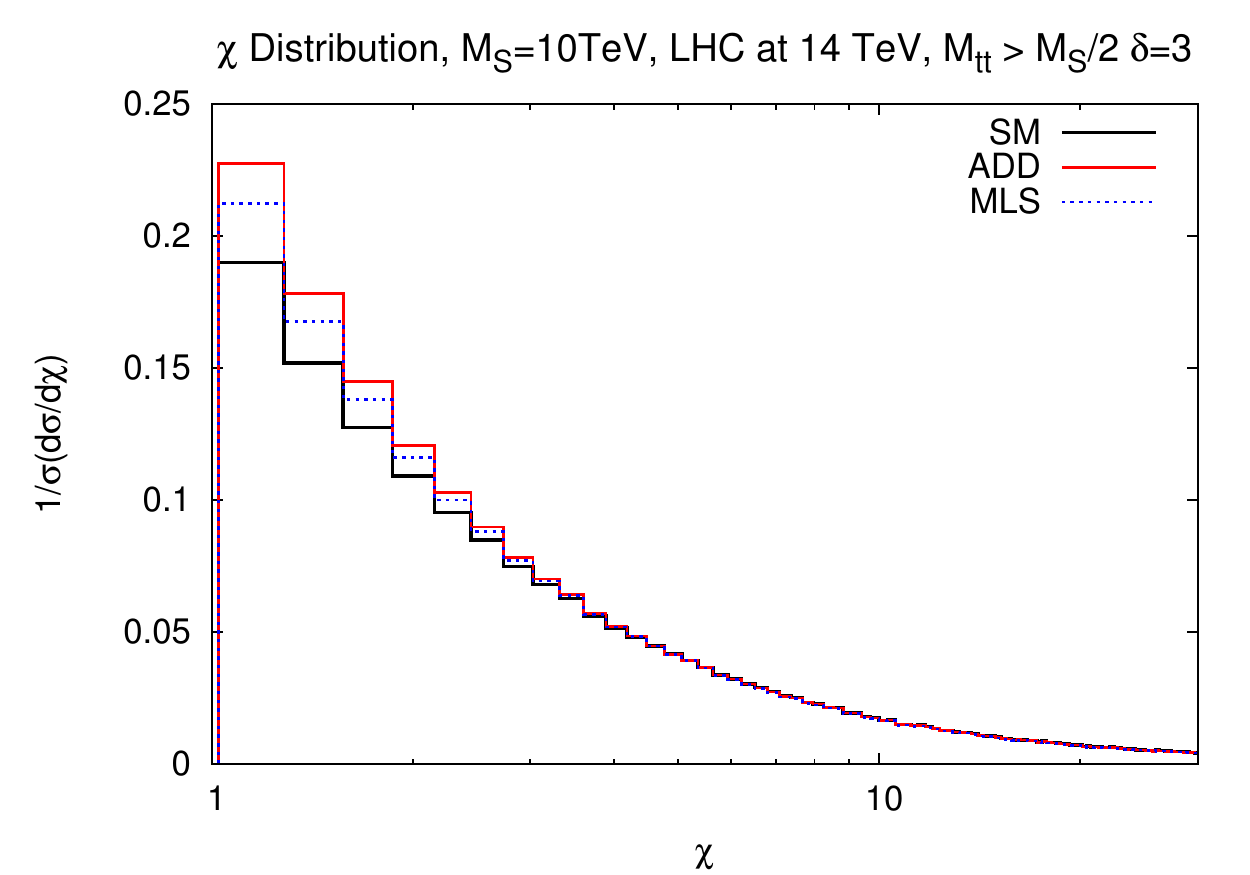}\\
\includegraphics[width=0.49\linewidth]{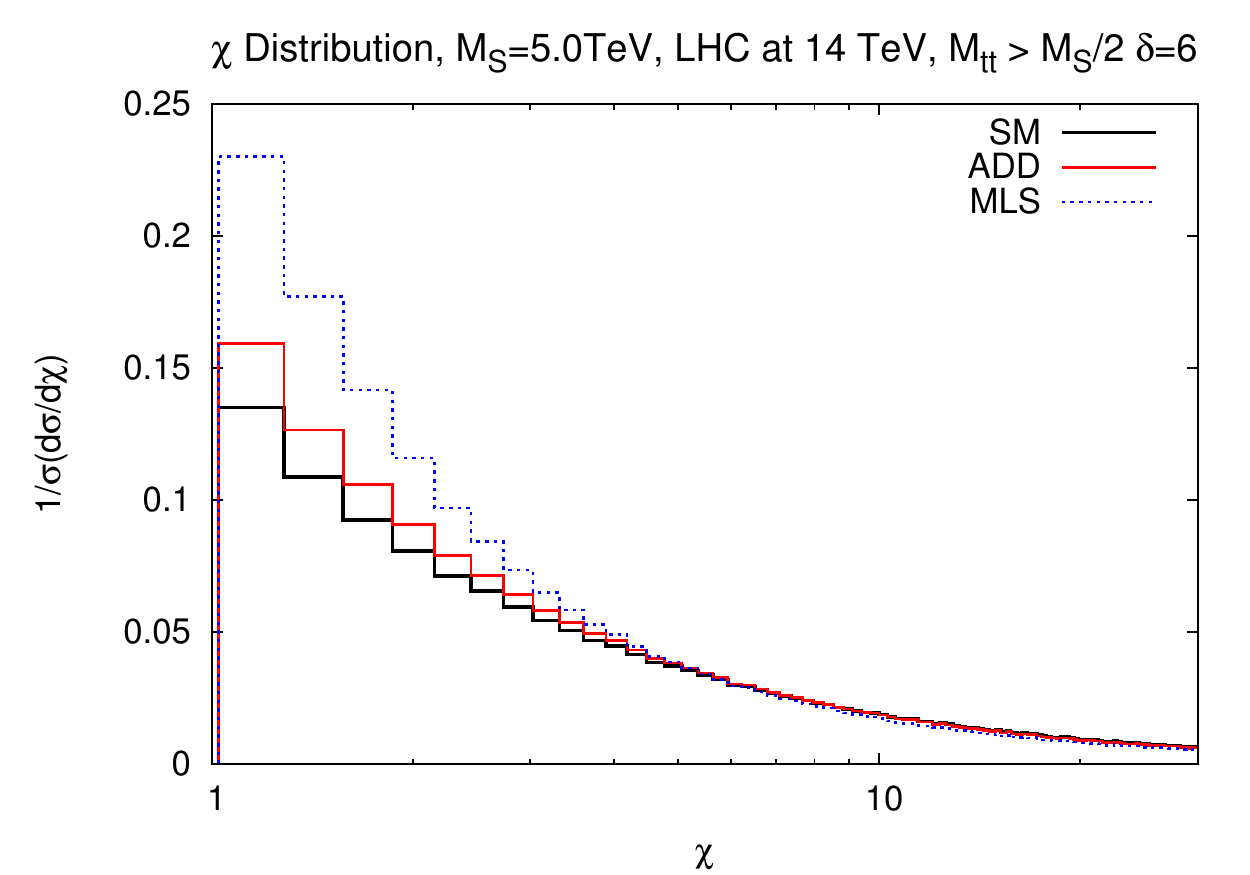}
\includegraphics[width=0.49\linewidth]{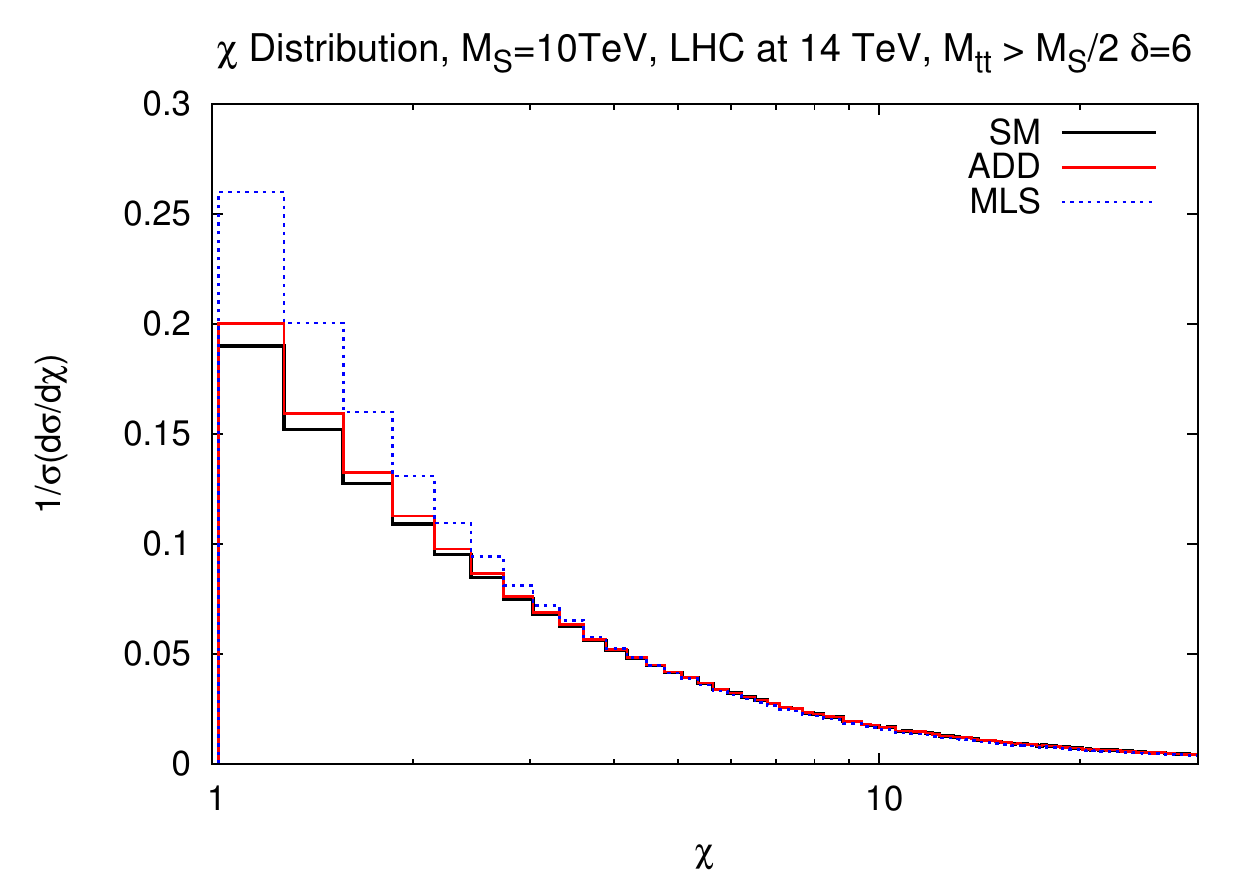}
\caption{ Differential cross sections w.r.t. the centrality ratio, $\chi$, at the LHC at 14 TeV for $M_{S}$~=~5 and 10 TeV for the SM, ADD and ADD-MLS for $\delta~=~3(6)$ XDs [\emph{upper}(\emph{lower})] normalised to 1 and compared to the SM expectation.}
\label{fig:lhc14chidists}
\end{center}
\end{figure}

\noindent 
Fig.~\ref{fig:lhc14chidists} shows the 
distributions normalised to 1 for $\delta~=~3,6$ XDs, implementing the 
tracking invariant mass cuts discussed in section~\ref{sub:mttdists}. Clearly, $\chi$ is a powerful variable at the LHC (at both energies)
as the ADD and MLS contribute significantly more in the low region (while this is not so at the Tevatron). The shoulders appearing in some of the distributions are an artefact of the invariant mass cuts and correspond to the $\chi$ value at the minimum invariant mass as $\cos\theta$ in the CM frame tends to 1. Additional cuts of $\chi< 4$ optimised to improve LHC reach were decided upon based on these. Although not addressed any further in this study, this variable may indeed be one of the best choices for directly setting bounds on signals induced by virtual gravitons, as applied in~\cite{Franceschini2011}. However, obtaining accurate angular information based on reconstructed $t\bar{t}$ final states is non-trivial.

\subsection{Total cross sections}
\label{sub:total}
\begin{figure}[!h]
\begin{center}
\includegraphics[width=0.49\linewidth]{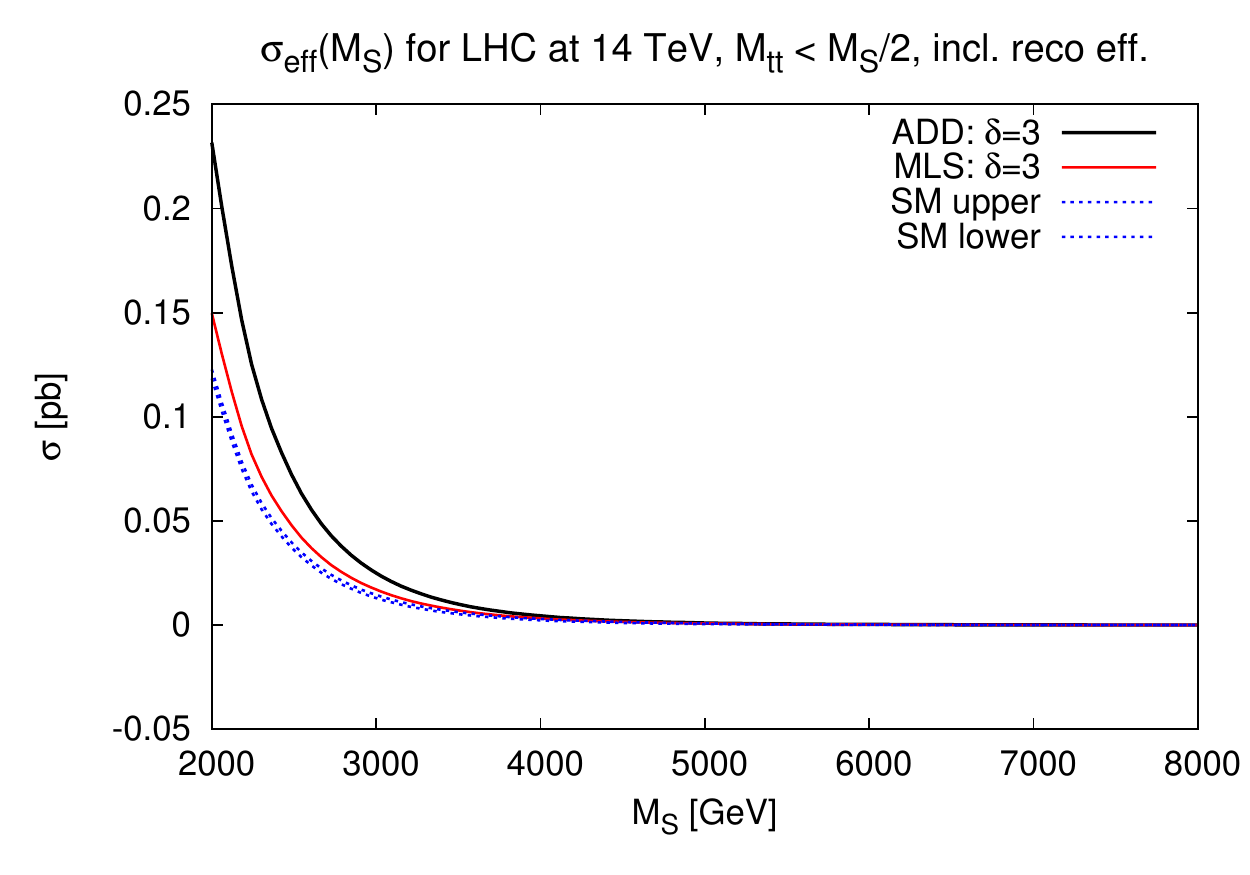}
\includegraphics[width=0.49\linewidth]{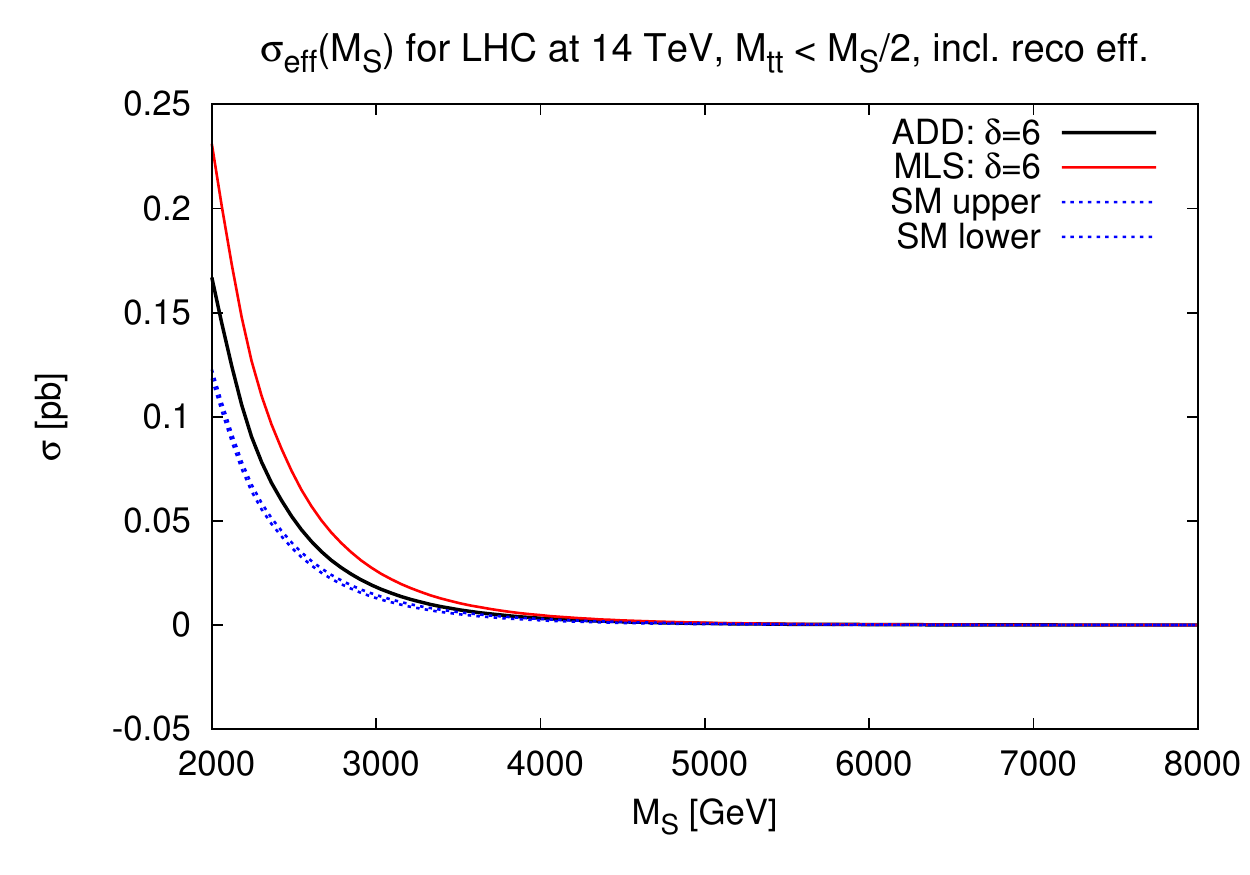}
\includegraphics[width=0.49\linewidth]{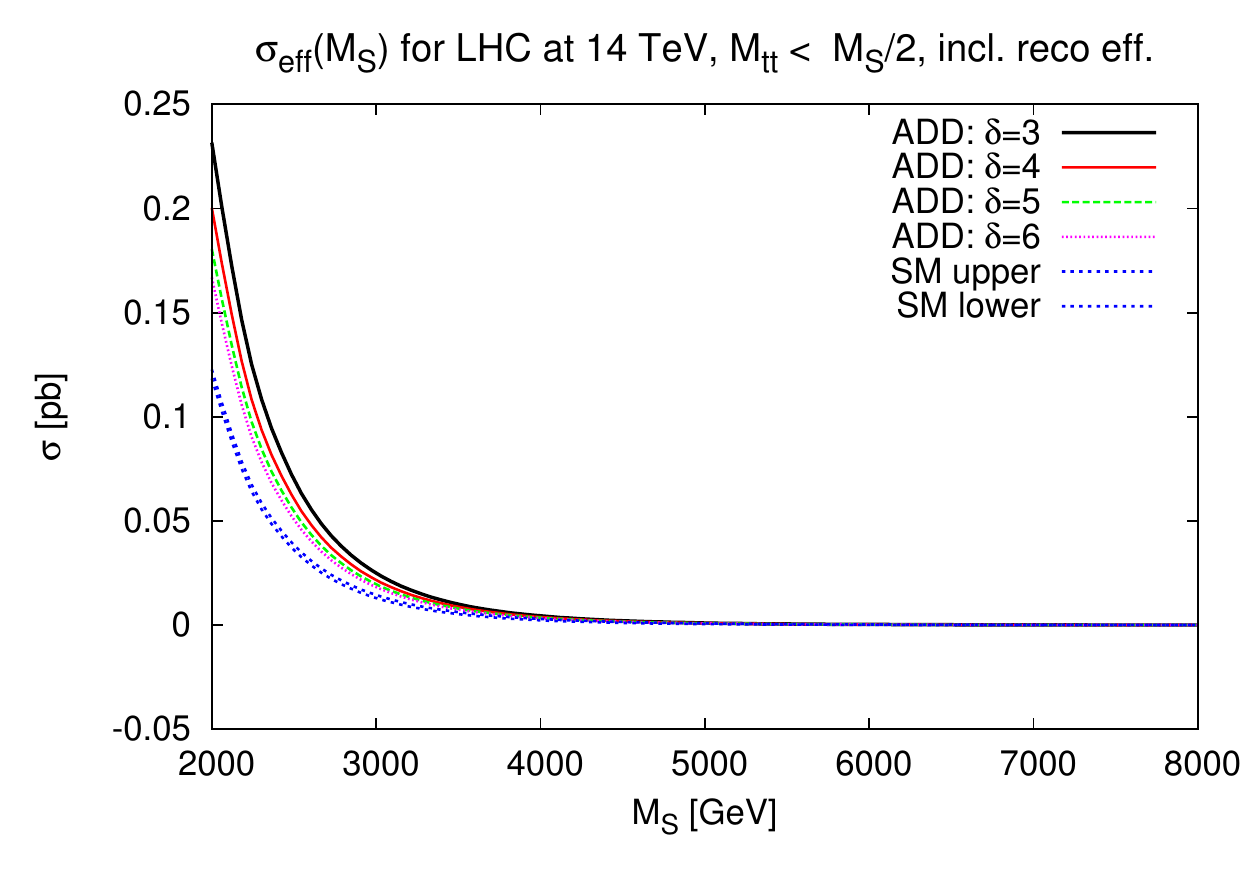}
\includegraphics[width=0.49\linewidth]{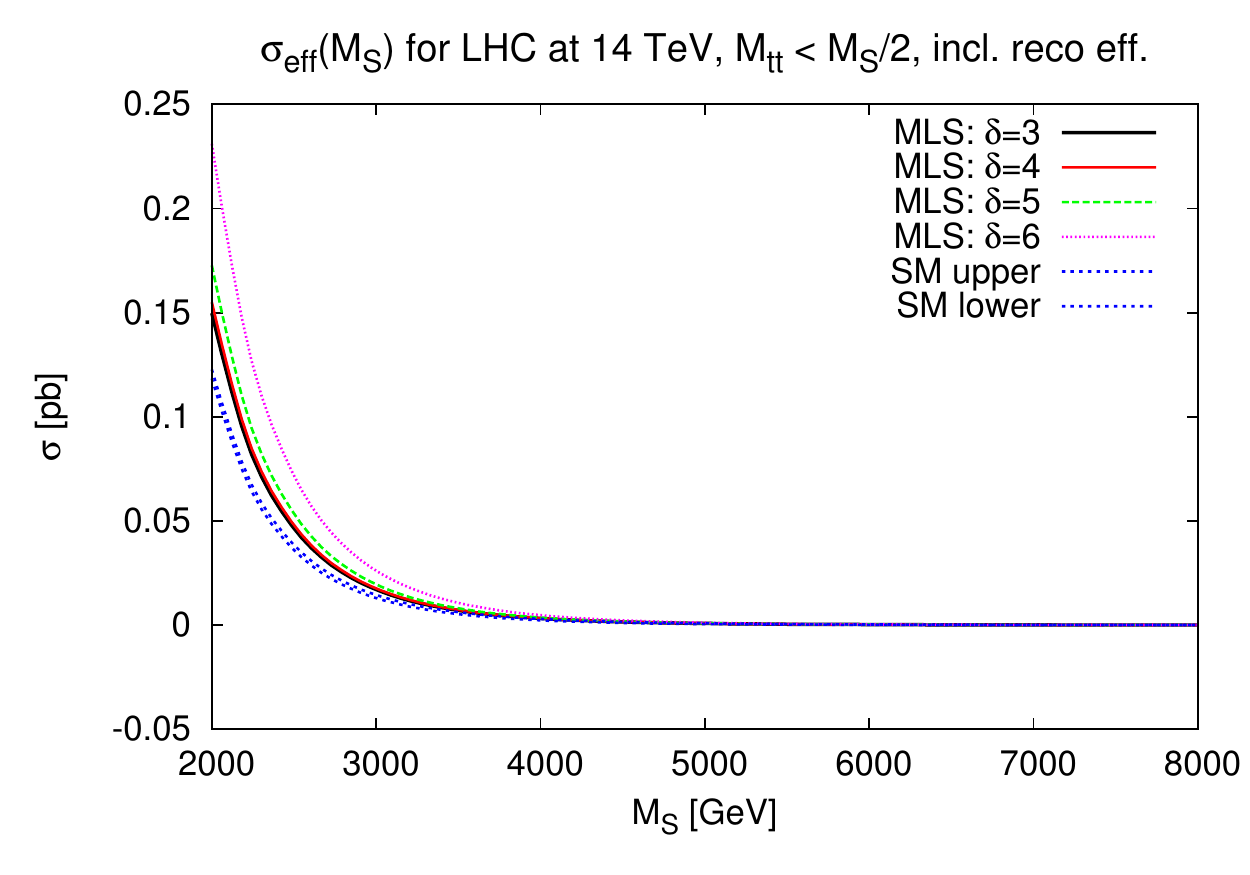}
\includegraphics[width=0.49\linewidth]{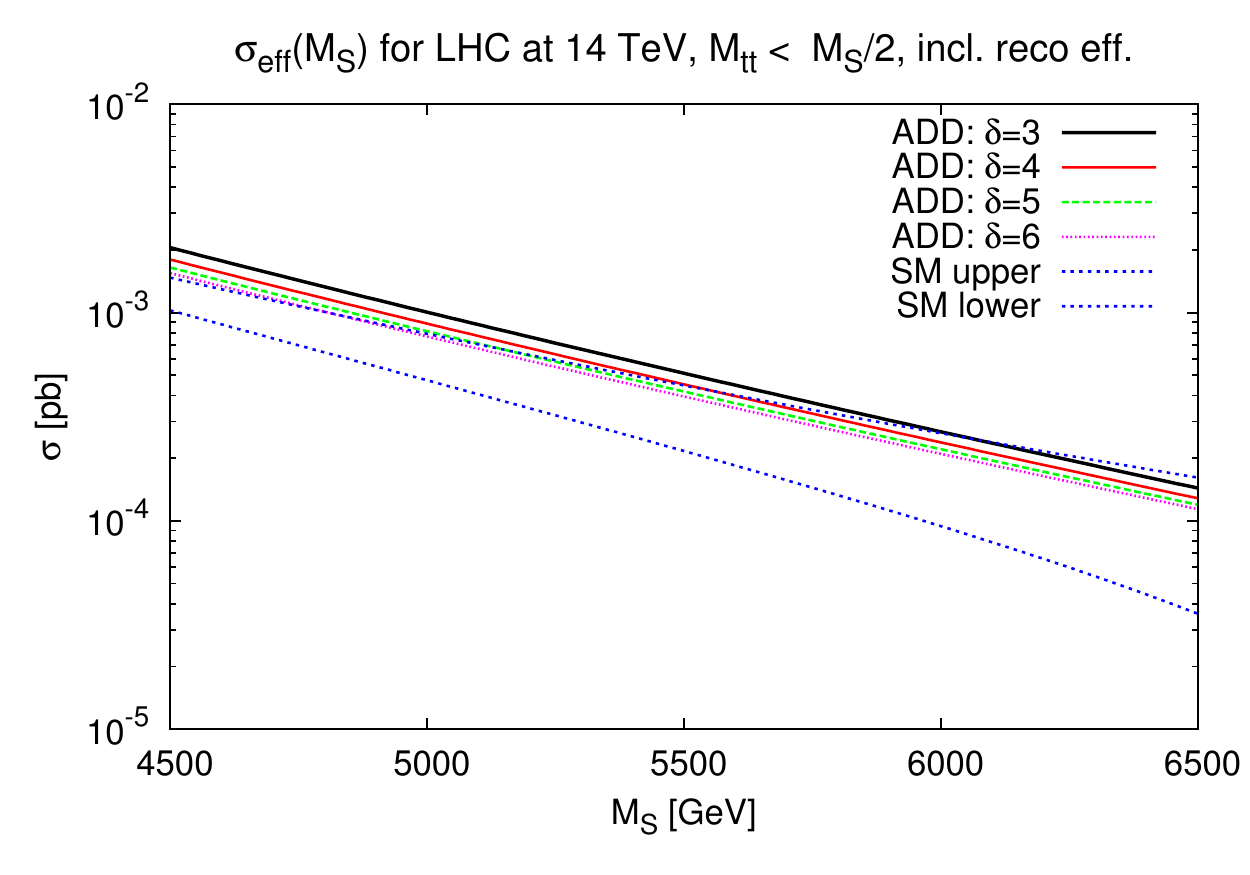}
\includegraphics[width=0.49\linewidth]{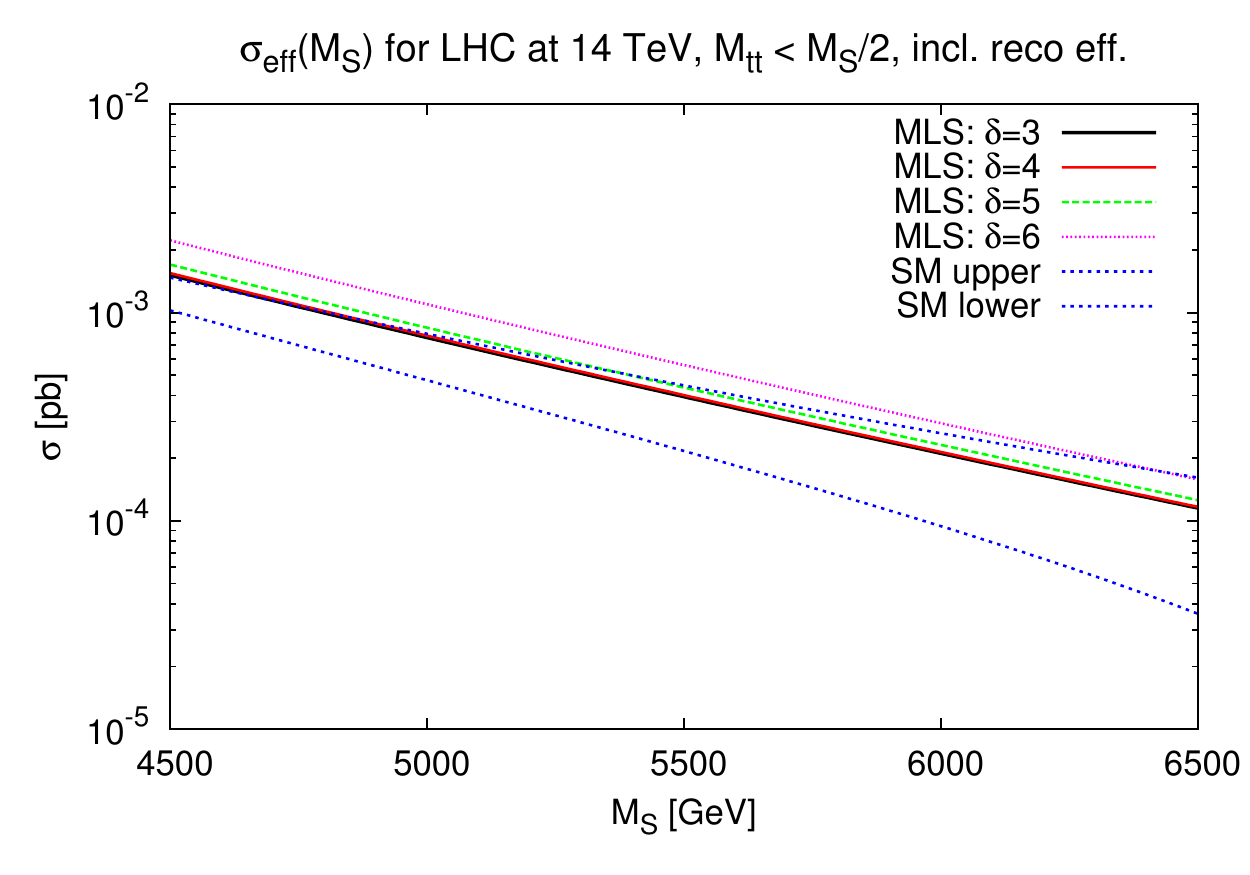}
\caption{ 
\emph{Upper}:    Total cross section for $pp\rightarrow t\bar{t}$ at the LHC with $\sqrt s=$~14 TeV as a function of $M_S$ for $\delta~=~3,6$ XDs comparing ADD 
and ADD-MLS.\newline
\emph{Middle}: Total cross section for $pp\rightarrow t\bar{t}$ at the LHC with $\sqrt s=$~14 TeV as a function of $M_S$ for $\delta~=~3-6$ XDs in ADD and ADD-MLS
separately.\newline
\emph{Lower}: Close-up on regions of interest in the above two plots.
Also plotted are the SM 95\% confidence level (CL) upper and lower bounds for 100 fb$^{-1}$ of integrated luminosity. 
Cuts on $M_{tt}$ and $\chi$ were enforced and an estimate of the $t\bar{t}$ reconstruction efficiency folded in, 
as described in the text.}
\label{fig:lhc14total}
\end{center}
\end{figure}

The plots of total cross section versus $M_{S}$ in Fig.~\ref{fig:lhc14total} show a similar trend to those presented 
in~\cite{Bhattacharyya:2004dy}. Based on the differential distributions discussed in the previous 
subsections, the tracking $M_{tt}>M_{S}/2$ and $\chi<4$ cuts were implemented. The 
differential cross sections were integrated up to $0.8M_{S}$ and compared to the 
total SM cross section integrated over the same range. In this way, the reach to set lower bounds on $M_{S}$ 
was determined for the case of non-observation of a deviation 
at 95\% CL for an assumed integrated luminosity of 100(5)[4] $\text{fb}^{-1}$ at the 
LHC(LHC)[Tevatron] with 14(7)[2] TeV. Simply plotting the total $t\bar{t}$ production cross section as a function of $M_{S}$ provided an idea of the potential to exclude regions of $M_{S}$ by comparing it to the upper and lower $2\sigma$ bounds for the SM prediction. The region of interest is where the signal becomes indistinguishable from the SM background for the assumed integrated luminosity.\\

\noindent
In Tab.~1 we summarise the rough bounds we obtain, after assuming a total 4(3)\%
reconstruction efficiency of the $t\bar t$ final state at 
the LHC(Tevatron)~\cite{deRoeck:942733,*Lukens:2003aq}, including
all possible decay channels. 

\begin{table}[h!]
\begin{center}
\begin{tabular}{cccc}
\hline
Collider & Luminosity &$M_S$ (ADD)& $M_S$ (ADD+MLS)\\
\hline
LHC at 14  TeV& 100 fb$^{-1}$ &5.4 TeV&5.1 TeV\\
LHC at 7  TeV& 5 fb$^{-1}$ &2.2 TeV&2 TeV\\
Tevatron at 2 TeV & 4 fb$^{-1}$ & $<$1 TeV&$<$0.8 TeV\\
\hline
\end{tabular}\caption{The 95\% CL limits on the reach to set bounds on $M_S$ in ADD and ADD-MLS at the three collider benchmarks considered. The efficiency
of reconstructing the $t\bar t$ final states in all possible decay channels is included and
cuts on $M_{tt}$ and $\chi$ were enforced, as described in the text. }
\label{tab:limits}
\end{center}
\end{table}

\subsection{Luminosity dependence}
\label{sub:lumi}
\begin{figure}[!h]
\begin{center}
\includegraphics[width=0.49\linewidth]{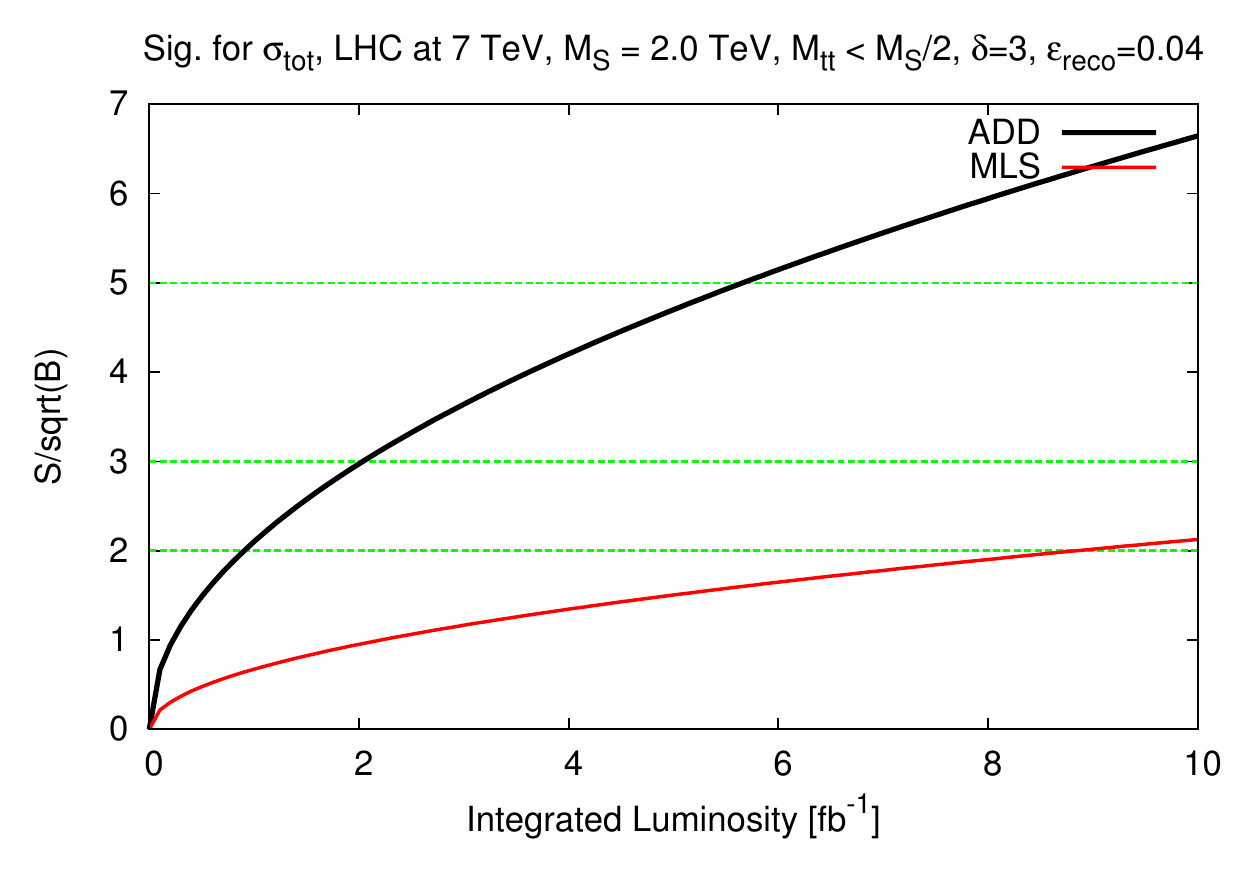}
\includegraphics[width=0.49\linewidth]{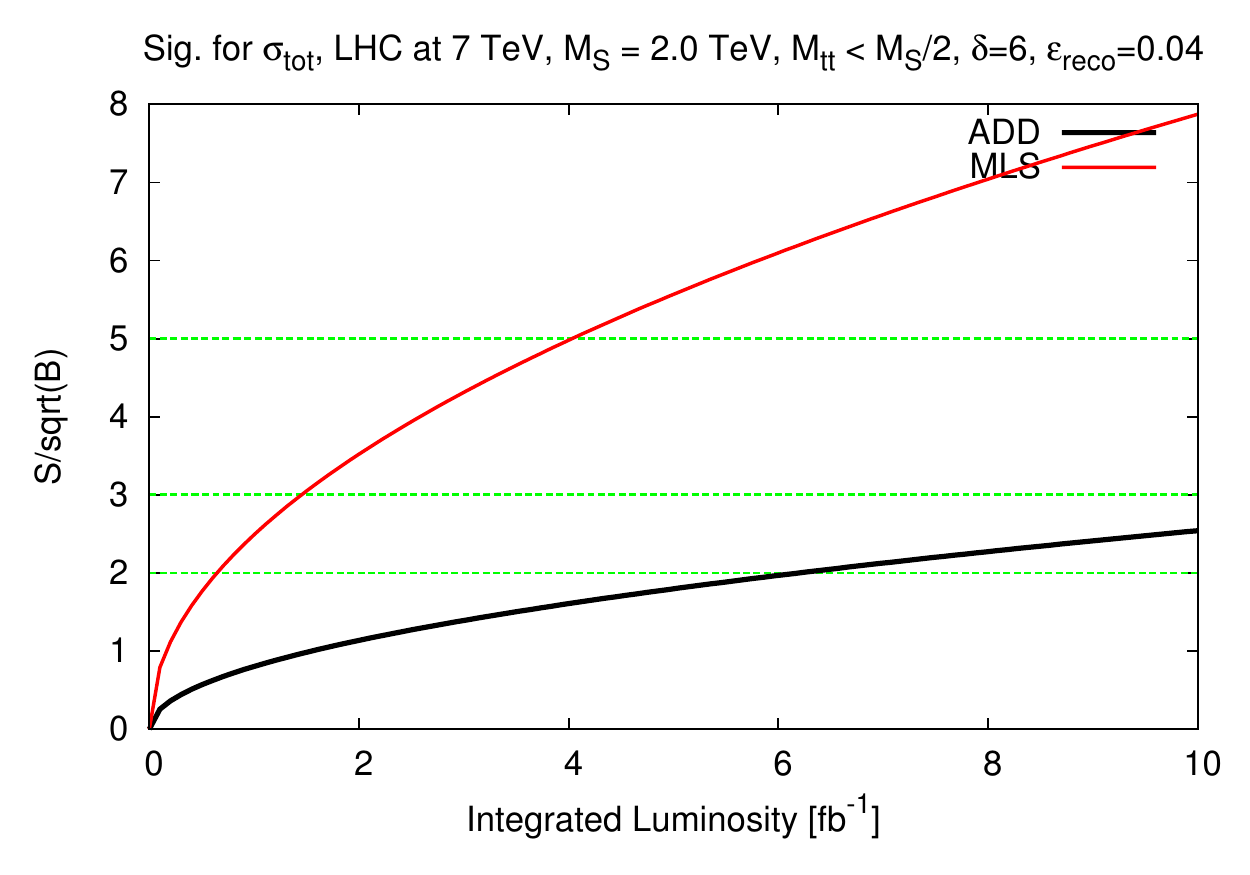}
\includegraphics[width=0.49\linewidth]{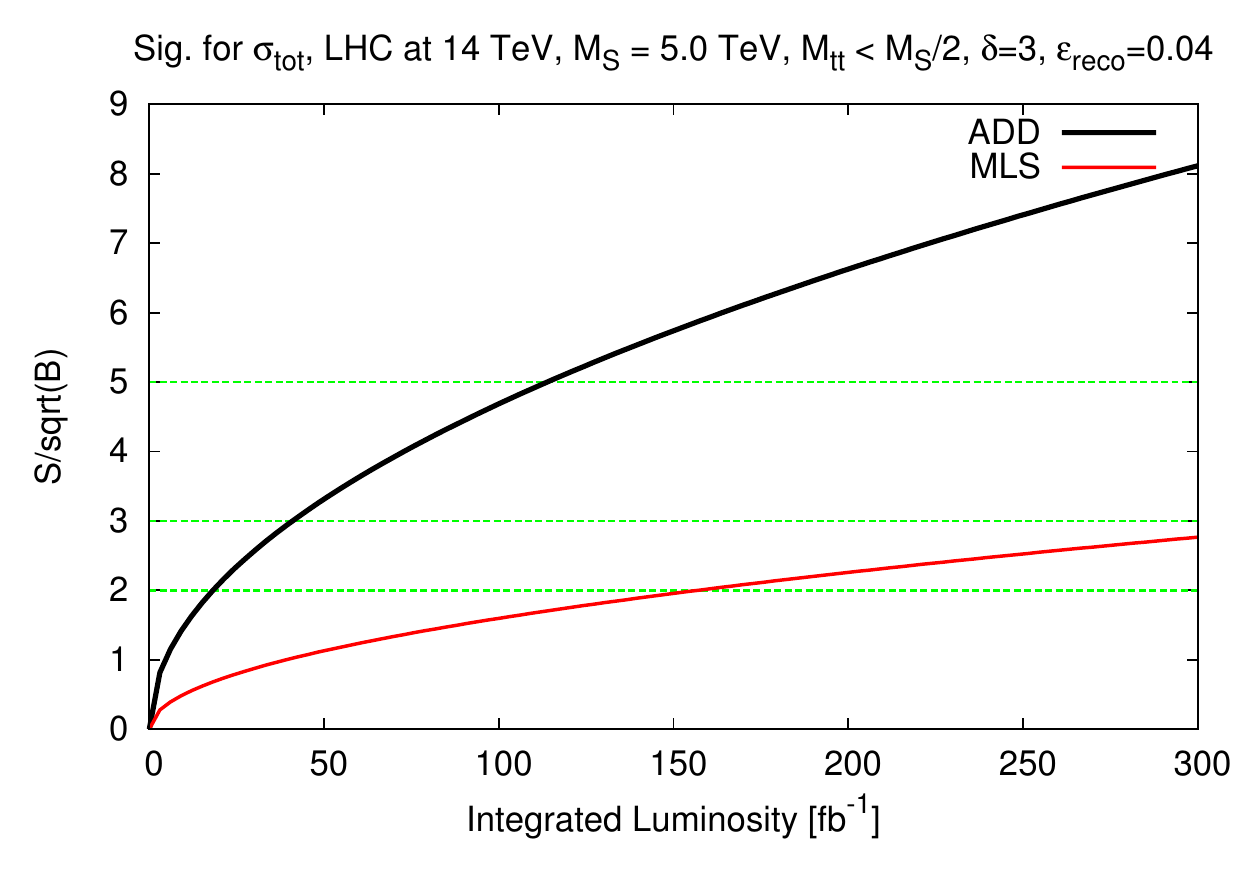}
\includegraphics[width=0.49\linewidth]{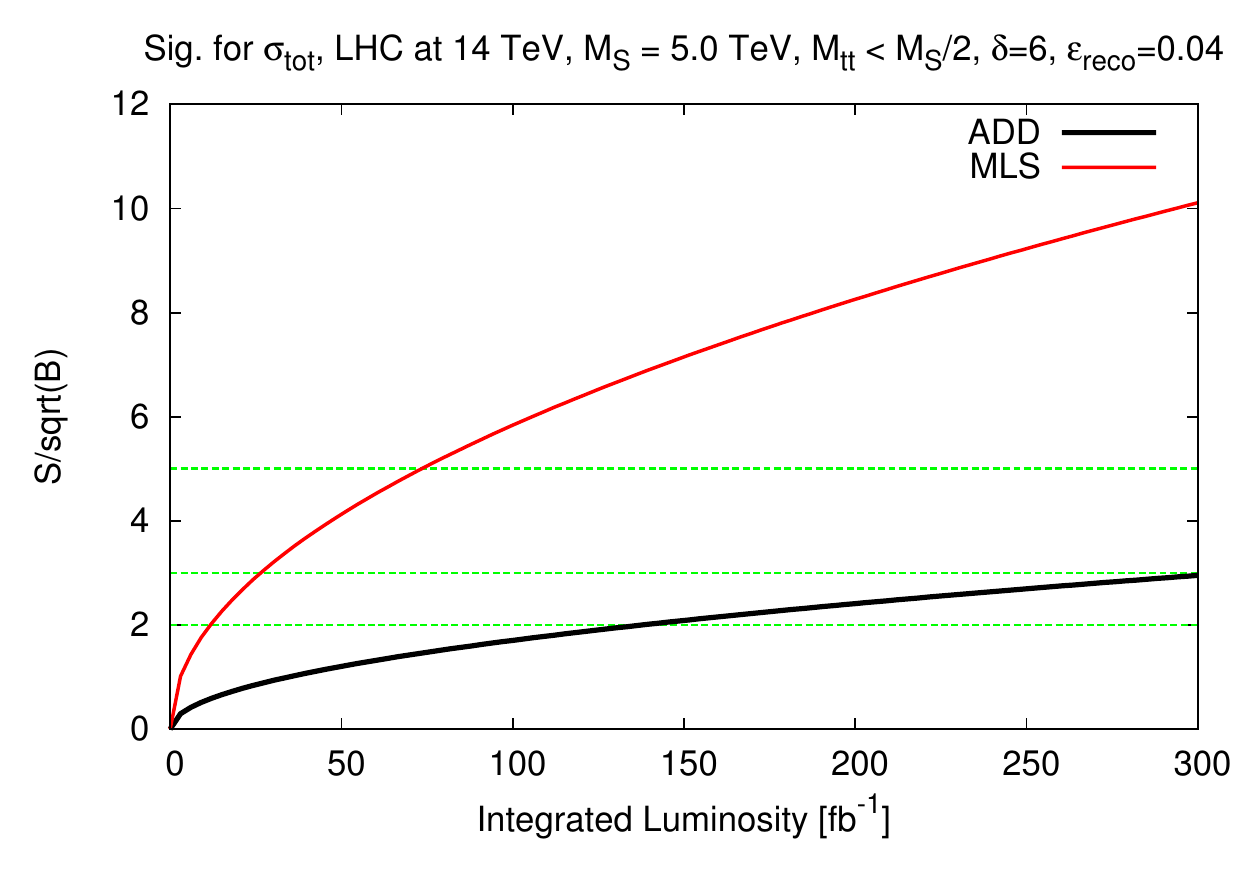}
\caption{ Significance, $\sigma=S/\sqrt{B}$, of the ADD and ADD-MLS with $M_{S}=2(5)$ TeV, $\delta~=~3,6$ as a function of the integrated luminosity for the LHC at 7(14) TeV [\emph{upper}(\emph{lower})]. This is calculated from the cross sections obtained imposing the tracking $M_{S}/2 < M_{tt} < 0.8M_{S}$ and $\chi < 4$ cuts. Lines of $\sigma=2,3,5$ are plotted for reference.}
\label{fig:lumi_lhc}
\end{center}
\end{figure}

\noindent
Finally, having assessed that the Tevatron scope is rather limited, we end the numerical analysis by 
taking some benchmark $M_{S}$ values for the LHC at 7 and 14 TeV and 
computing the corresponding integrated cross sections in order to determine the potential significance, $S/\sqrt{B}$, 
that could be obtained at the LHC collider as a function of the integrated luminosity, where $S$ and $B$ represent the number of 
signal and background events, respectively. 
The aforementioned $M_{S}/2<M_{tt}<0.8M_{S}$ and $\chi<4$ cuts were used. 
The significances are shown in Fig.~\ref{fig:lumi_lhc} for values of $M_{S}$ 
that should be visible at the two LHC configurations, according to the results of subsec.~\ref{sub:total}. 
These results are consistent with those in the previous subsection for the assumed values of integrated luminosity and show 
in perspective that such models are further testable and clearly distinguishable from each other 
not only by the LHC at 14 TeV but also, to an understandably more limited extent, by the 
LHC at 7 TeV, with increasing data. 
\clearpage

\section{Conclusions}
\label{sect:summa}
In summary, the $t\bar{t}$ channel at the LHC shows some promise in being able to access and set bounds on 
the parameters of the ADD model with and without the MLS extension, specifically, when
$M_{S}\leq 5$ TeV for $\delta~=~3-6$, assuming design energy and luminosity. Compared to a similar study 
conducted in the DY channel, the limited efficiency in reconstructing $t\bar t$ pairs 
(with respect to $e^+e^-$ and $\mu^+\mu^-$ pairs) means that the bounds obtained in this study may not 
quite surpass those suggested in~\cite{Bhattacharyya:2004dy}. The early data LHC setup has a more limited
but still relevant reach ($M_{S}\simeq 2$ TeV for $\delta~=~3-6$) 
while the Tevatron always performs badly in the $t\bar t$ channel, 
mainly due to the lower energy and dominance of the $q\bar{q}$ initial 
state, which has no ADD/SM interference term. 
Notwithstanding, it is clear that this channel can prove useful in probing such XD models as a complement 
to other, perhaps `cleaner' channels, such as DY, not only in measurements of the total cross section but 
also by considering differential distributions such as the invariant mass of the top quark pair and its 
centrality ratio. We have in fact shown that these distributions could reveal significant deviations 
from the SM predictions for values of $M_{S}$ up to around 5(2.5) TeV for the LHC at 14(7) TeV and could merit further inquiry. \\

\noindent
In the MLS extension of the ADD model, an alternative method for regulating the UV divergences of the 
virtual KK graviton sum has been explored. We have shown that it tames some undesirable behaviour of the 
ADD cross section using a natural prescription inspired by String theory considerations as one approaches the effective cutoff 
of such models, $M_{S}$, thereby avoiding any potential pathological overestimate of the cross sections involved. 
In the context of collider phenomenology, this scenario would have the effect of lowering the reach of 
the LHC to set bounds on or extract $M_{S}$ as a consequence of its regulating properties.\\

\noindent 
Finally, it is also worth remembering here that our results are based on tree-level calculations,
as NLO results in the ADD model (and modifications thereof) are not available yet. Our exercises should therefore eventually 
be repeated in presence of higher order effects, though we do not expect the gross features obtained
here to change dramatically.

\section*{Acknowledgments}
This work is financed in part by the NExT Institute and STFC (Swindon, UK). We thank K. Sridhar and S. Biswal for their collaboration during the initial stages of this work. KM thanks the Tata Institute for Fundamental Research (Mumbai, India) and the University of Southampton (through an International Fund grant) for travel support
during a visit to India, where part of the work took place.

\clearpage
\bibliography{references}

\end{document}